\documentclass[journal]{IEEEtran}
\ifCLASSINFOpdf
\else
\fi

\usepackage{graphicx}
\usepackage{amsmath, amssymb}
\usepackage{algorithm}
\usepackage{algpseudocode}
\usepackage{booktabs}
\usepackage{tablefootnote}
\usepackage{cite}
\usepackage[caption=false,font=footnotesize]{subfig}

\begin{document}
\title{A Game-Theoretic Decentralized Real-Time Control of Electric Vehicle Charging Stations - Part II: Numerical Simulations}

\author{Riccardo Ramaschi,
\IEEEmembership{Student Member, IEEE}, Mario Paolone, \IEEEmembership{Fellow, IEEE}, Sonia Leva, \IEEEmembership{Senior Member, IEEE}
\thanks{R. Ramaschi and S. Leva are with the Department of Energy, Politecnico di Milano, Milan, Italy. M. Paolone is with the Distributed Electrical Systems Laboratory, Swiss Federal Institute of Technology, Lausanne, Switzerland.}
}

\markboth{IEEE Transaction on Smart Grids}%
{Shell \MakeLowercase{\textit{et al.}}: Bare Demo of IEEEtran.cls for IEEE Journals}

\maketitle

\begin{abstract}
In the first part of this two-part paper a game-theoretic decentralized real-time control is proposed in the context of Electric Vehicle (EV) Charging Station (CS). This method, relying on a Stackelberg Game-based Alternating Direction of Multipliers (SG-ADMM), intends to steer the EVs' individual objectives towards the CS optimum by means of an incentive design mechanism, while controlling the EV power dispatch in a distributed manner. We integrate SG-ADMM in a hierachical multi-layered Energy Management System (EMS) as the real-time control algorithm, formulating the two-layer approach so that the SG leader (i.e., the CS), holding commitment power, trades off the available power with the incentives to the EVs, and the SG followers (i.e., the EVs) optimizes their charging curve in response to the leader decision. In this second part, we demonstrate the applicability of SG-ADMM as a incentive design mechanism inside an EVCS EMS, testing it in a large-scale EVCS. We benchmark this method with a decentralized (ADMM-based), a centralized and a uncontrolled approach, showing that our method exploits EV-level flexibility in a cost-effective, fair and computationally efficient manner.
\end{abstract}

\begin{IEEEkeywords}
Electric Vehicle Charging Station, Real-time Control, Game Theory, Alternating Direction Method of Multipliers, Incentive Design
\end{IEEEkeywords}

\section{Introduction}

\begin{figure}[t]
    \centering
    \includegraphics[width=\linewidth]{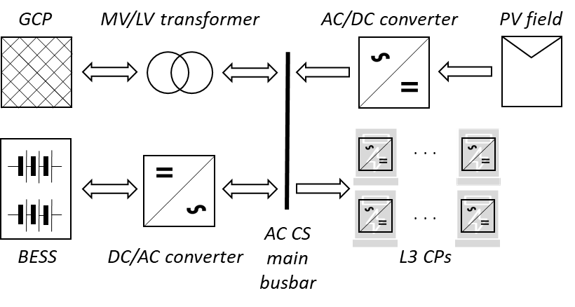}
    \caption{Charging station layout: components and configuration. The image is adapted from \cite{pt} with authors' permission.}
    \label{charging_station}
\end{figure}

In the first part of this two-part paper, we proposed to apply SG-ADMM to an Electric Vehicle (EV) Charging Station (CS), where the CS is the central controller anticipating the EVs' reaction by designing ad hoc incentives and the EVs are the followers optimizing accordingly their power in a distributed manner. In order to do that, we proposed some modifications to the original SG-ADMM so that the incentive provision is traded off against the available charging power, that is the ADMM coupling constraint, by means of a bisection method. The full integration of this tweaked SG-ADMM version is presented in the first part, while in this second part we benchmark and evaluate the performance of the proposed method through numerical simulations. The algorithm is tested on an EVCS whose structure is similar to the one proposed in \cite{pt}, that is a unidirectional Level 3 (L3) CS with grid, Photovoltaic (PV) and Battery Energy Storage System (BESS) connection as shown in Figure \ref{charging_station}. 

Since the proposed Energy Management System (EMS) formulation consists of three layers, this second part details each layer inputs, focusing on the PV and EV forecasting pipeline. Besides the numerical evaluation of a state-of-the-art incentive design mechanism in a novel context, this second part contributes by developing a comprehensive EV forecast framework that enables to predict EV demand throughout all the EMS's granularities.

Part I mainly contributed by:
\begin{itemize}
    \item Reviewing the literature on the use of SG-ADMM for single-leader multi-followers non-cooperative games, identifing a contextual research gap on the deployment of the method.
    \item Formulating a comprehensive hierarchical multi-layered EMS for EVCS control.
    \item Framing the real-time control and the corresponding incentive mechanism as a SG-ADMM algorithm.
    \item Accomodating the SG-ADMM method to the specific case studies, especially by developing the leader incentive update method that trades off the available power with the incentive provision.
\end{itemize}

On top of these contributions, Part II contributes by:
\begin{itemize}
    \item Instantiating the employed case study and the corresponding forecasting frameworks.
    \item Developing a novel prediction pipeline for EV power curve forecast, integrated in a wider EV demand prediction module.
    \item Benchmarking the proposed method with an hard-to-beat centralized approach, a not-incentivizing decentralized approach and a baseline uncontrolled approach.
    \item Evaluating the applicability of SG-ADMM in terms of power dispatch, computational time, economic performance, battery wear, charging time and fairness.
\end{itemize}

The remainder of this second part is organized as follows. Section \ref{recall} recalls the EMS formulation and Section \ref{numerical} contextualizes it in a specific case study, defining the CS parameters and forecasting modules. Section \ref{benchmarks} introduces the SG-ADMM benchmarks and the shared real-time dispatch algorithm. Finally, Section \ref{results} presents the results and the conclusions are drawn in Section \ref{concl}.

\section{Recall of the EMS formulation} \label{recall}
For the sake of self consistency of this second part, we briefly recall the EMS structure proposed in Part I. The proposed EMS consists in three layers, i.e. a chance-constrained Day-Ahead (DA) optimization defining the Dispatch Plan (DP), a Intra-Day (ID) refinement of the DP according to updated information and the power budget concept \cite{Power_budget}, and a real-time control and incentive mechanism based on SG-ADMM. Here are briefly reported their operations:
\begin{itemize}
    \item The day-ahead layer is framed as a chance-constrained stochastic problem that aims at submitting the Day-Ahead Market (DAM) transaction considering possible speculation on the Balancing Market (BM).
    \item The intra-day layer aims at refining the day-ahead power schedule considering the updated forecast information, setting the grid power budget and the BESS setpoint. This refinement is performed in two different ways for short-term and medium-term time spans according to a sliding window approach.
    \item The real-time layer consists on the nested ADMM (inner) and SG (outer) iterations: the SG leader (CS) updates the coupling constraint and the incentives, while the SG followers (EVs) optimize their power demand accordingly in a distributed manner.
\end{itemize}

\section{Numerical Simulations} \label{numerical}
The proposed model is applied in a specific case study, that is shown in Section \ref{case} together with the hyperparameters chosen for the model fine-tuning. The inputs of the model, i.e. the forecasts, are explained in Section \ref{for}.
\subsection{Case study} \label{case}
\begin{figure}[t]
    \centering
    \includegraphics[width=1\linewidth]{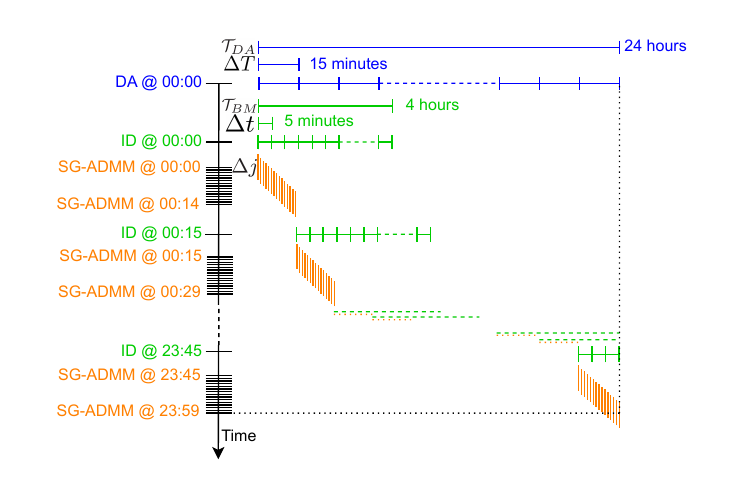}
    \caption{Time frames of the proposed three-layer EMS: horizon and granularity of a daily simulation}
    \label{frames}
\end{figure}

We present in Table \ref{case_study} the parameters chosen for the CS characterization and the model time frames (the models sequence during a daily simulation is shown in Figure \ref{frames}).

\begin{table}[t] 
    \caption{CS parameters and model time frames}  
    \centering
    \begin{tabular}{c c c c c c} 
        \toprule
        \textbf{Symbol} & \textbf{Value} & \textbf{Symbol} & \textbf{Value} & \textbf{Symbol} & \textbf{Value} \\
        \midrule 
        $\eta_{pv}$ & 0.98 & $P^{peak}_{PV}$ & 500 kW & $\eta_{ch}, \eta_{dh}$ & 0.95\\
        $\eta_{inv}$ & 0.98 & $C_{B}$ & 506.7 kWh & $p_{kWh}$ & 115 \$/kWh\\
        $c_{rate}$ & 1 & $SoC_{min}$ & 10\% & $SoC_{max}$ & 90\%\\ 
        $\eta_{tr}$ & 0.99 & $P_{GC}$ & 954.5 kW & $\eta_{cp}$ & 0.95\\
        $n_{CC}$ & 10 & $P_{CC}$ & 172.5 kW & $n_{CP}$ & 2\\
        $\mathcal{T}_{DA}$ & 24 h & $\mathcal{T}_{BM}$ & 4 h & $\Delta T$ & 15'\\
        $\Delta t$ & 5' & $\Delta j$ & 1' & $D_{B,EoL}$ & 80 \%\\
        \bottomrule
    \end{tabular}
    \label{case_study}
\end{table}

The CS is located in Lausanne - Switzerland - and the BESS capacity and Grid Connection Power (GCP) are sized accordingly using the model developed in \cite{pt} applied to a Swiss case study. DAM prices are taken from \cite{swiss_da}, EV data is taken and scaled from \cite{desl_db}, PV data is taken from the EPFL Distributed Electrical Systems Laboratory (DESL) in Lausanne \cite{desl}, GCP connection costs are taken from \cite{romandie}. Please refer to \cite{pt} for the model and other parameters.

For the daily simulations, we used the following databases:
\begin{itemize}
    \item DAM prices ($\mathcal{R}_{DAM}$) from \cite{swiss_da},
    \item BM prices ($\mathcal{R}_{BM}$) from \cite{swissgrid},
    \item EV charging rate ($\mathcal{C}$) from \cite{tesla_prices},
    \item EV charging database from \cite{desl_db},
    \item PV power production from \cite{desl} measurements.
\end{itemize}

In Table \ref{hyperparam} the hyperparameters used in the model are presented. The majority of the parameters have been chosen after a fine tuning that is out of the scope of the paper, although the underlying choices are justified in Part I. Instead, $\rho=10$, $\epsilon_{abs}=0.0001$, $\epsilon_{rel}=0.01$, $\varepsilon=0.001$, $\tau_{\rho}=2$ and $\mu=10$ as suggested in \cite{Boyd_ADMM}.

\begin{table}[] 
    \caption{Model hyperparameters}  
    \centering
    \begin{tabular}{c c c c c c} 
        \toprule
        \textbf{Symbol} & \textbf{Value} & \textbf{Symbol} & \textbf{Value} & \textbf{Symbol} & \textbf{Value} \\
        \midrule 
        $f_s$ & 0.8 & $b$ & 1 & $c$ & 0.01\\
        $\tau$ & 0.2 & $a$ & 0.04 & $\alpha$ & 10\\
        $\beta$ & 0.01 & $\gamma$ & 10 & $\rho$ & 10\\
        $\delta$ & 0.04 &$\epsilon_{abs}$ & 0.0001 & $\epsilon_{rel}$ & 0.01\\
        $\varepsilon$ & 0.001 & $\tau_{\rho}$ & 2 & $\mu$ & 10\\
        \bottomrule
    \end{tabular}
    \label{hyperparam}
\end{table}
\subsection{Forecast modules} \label{for}
The EMS is equipped with PV production, EV demand and an imbalance price forecast modules.
\subsubsection{PV forecast}
\begin{figure}
    \centering
    \includegraphics[width=\linewidth]{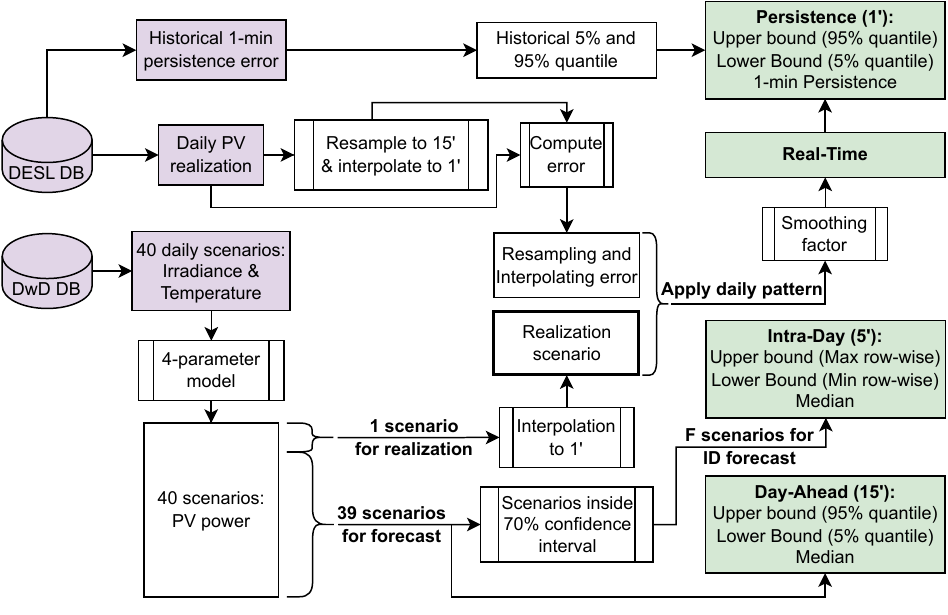}
    \caption{PV forecast and realization pipeline. It generates DA, ID and 1-min persistence forecast and a real-time realization.}
    \label{pv_pipeline}
\end{figure}

The EMS relies on three different forecaster according to the horizon for the PV production prediction. In this work, we adopt a unified PV-forecasting framework that consistently generates forecasts and realization from the same scenario-based process. This simplifying assumption has been undertaken to preserve coherence across all stages of the forecasting chain and to ensure full consistency between the forecasts and the realization against which they are assessed, considering that PV forecast is outside of our paper scope. We present in Figure \ref{pv_pipeline} the proposed PV pipeline.

We start by using the data coming from DwD \cite{dwd} providing 40 scenarios for irradiance and temperature for the chosen site (EPFL DESL in Lausanne). We managed this data by mapping the 40 scenarios of irradiance and temperature into 40 scenarios of PV production through a four-parameter physical model (i.e., PV peak power of 500 kW, azimuth of 180° south, tilt angle of 25° and coordinates of 46°31' N and 6°33' E). From here, we remove one scenario (avoiding information leakage) to generate the realization scenario, originally obtained interpolating to 1 minute the originally 1-hour granular DwD data. Thus, we remain with 39 scenario; considering a 90\% confidence level, we compute the 5\% and 95\% quantiles on top of the median (corresponding to $\hat{P}^{\downarrow}_{PV,T}$,  $\hat{P}^{\uparrow}_{PV,T}$ and $\hat{P}_{PV,T}$ of Part I, respectively). The DA problem is therefore fed with $\hat{P}^{\downarrow}_{PV,T}$, $\hat{P}^{\uparrow}_{PV,T}$ and $\hat{P}_{PV,T}$ with a 24-hour horizon and a 15-minutes granularity (interpolated from the 1-hour original granularity). 

To account for the expected refinement of the ID forecast, we limit the number of scenarios from 39 to $F$ by selecting those scenarios lying inside the 70\% confidence interval. From these scenario, we find the median, corresponding to $\hat{P}_{PV,t}$ of Part I. For the ID upper and lower bound, we create them by taking, for each hour, the maximum value across all scenarios for the upper bound (maximum row-wise) and the minimum value across all scenarios for the lower bound (minimum row-wise). In this way we find $\hat{P}^{\downarrow}_{PV,t}$ and $\hat{P}^{\uparrow}_{PV,t}$ of Part I. The ID problem is therefore fed with $\hat{P}^{\downarrow}_{PV,t}$, $\hat{P}^{\uparrow}_{PV,t}$ and $\hat{P}_{PV,t}$ with a 4-hour horizon and a 5-minutes granularity (interpolated from the 1-hour original granularity).

To create more realistic PV realization, the original 1-minute granular realization scenario is post-processed as follows. First, the actual daily PV realization is obtained from the EPFL DESL measurements \cite{desl} (referred to as reference). It is resampled to 15 minutes and interpolated to 1 minute applying the same value to the whole 15-minutes range (resampled-interpolated). Then, the per unit deviation between the reference and the resampled-interpolated realizations is taken and applied to the realization scenario. In this way, we are applying the actual daily pattern to an artificially created realization scenario. To avoid inconsistent behavior, we scale the error by a smoothing factor (3). In this way we find the real-time PV realization, corresponding to $P_{PV,j}$ of Part I.

Finally, the real-time control, regardless from the method, is performed through a 1-minute naive persistence for what concerns the expected PV production. We use $P_{PV,j-1}$ as the 1-minute persistence forecast, while the upper and lower bound are obtained as follows. First, from the DESL measurements \cite{desl} we obtain the historical 1-minute persistence error and we get the 5\% and the 95\% quantile values. We directly apply to the 1-minute persistence these values to get the upper ($P^{\uparrow}_{PV,j-1}$) and lower bound ($P^{\downarrow}_{PV,j-1}$). This concludes the PV forecast and realization pipeline presented in Figure \ref{pv_pipeline}. 

Instead, for the upsampling performed during the intraday refinement for the short-term horizon (15 minutes), a model called Robust Persistence (RP) is derived from \cite{Nowcasting_SM} and applied with a 1-minute granularity at every intraday model occurrence:
\begin{equation}
    P_{j:j+15'}=P_{j-1} \cdot \frac {A_{j}}{A_{j-1}}
\end{equation}
Where $P_{j-1}$ is the measurement at the previous time step, $P_{j:j+15'}$ is the forecast over 15 minutes and $A_j$ is the sun elevation at time $j$. An error analysis allowed to chance-constrain the PV realization in time step $j$ through a 95\% confidence interval error analysis. Therefore, we generate the median, the 2.5\% and the 97.5\% quantiles for the RP forecast, i.e. $\hat{P}_{PV,j}$, $\hat{P}^{\downarrow}_{PV,j}$ and $\hat{P}^{\uparrow}_{PV,j}$ from Part I.

\subsubsection{EV forecast}
The EV forecast module is divided in two main parts. Both parts rely on the database \cite{desl_db}. The DA prediction method relies on a Gaussian Mixture Model (GMM) based method proposed in \cite{Grid_aware}. We assume that the module receives the arrival and departure time information of the arriving EVs on the intraday horizon $\mathcal{T}_{BM}$, i.e. 4 hours. Therefore, the second part is an EV booking system with a 5-minutes granularity that allows to extract easily the maximum EV demand, $P^{max}_{EV,t}$, and the expected EV demand, $\hat{P}_{EV,t}$, (both needed as per the EMS formulation of Part I) via a comprehensive method described below.

\begin{figure}[t]
    \centering
    \includegraphics[width=\linewidth]{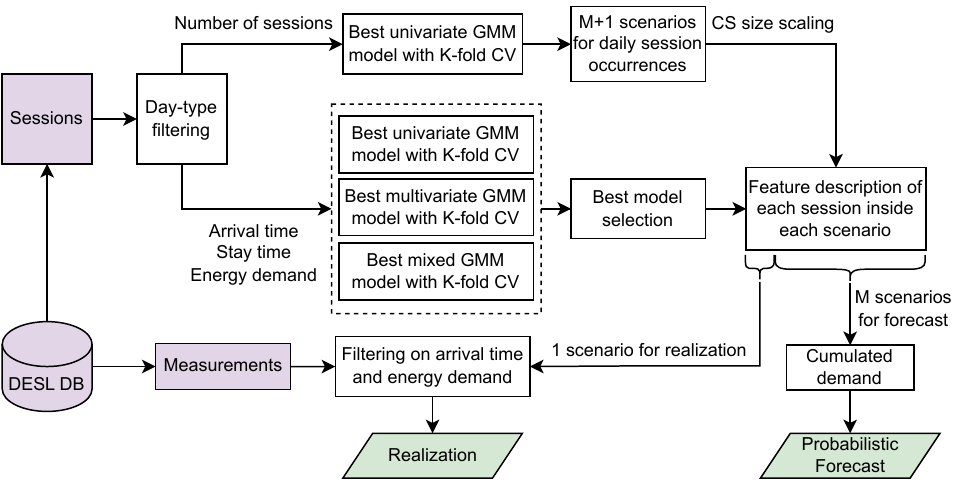}
    \caption{Proposed method for EV scenario generation and forecasting using GMM (similar approach to the one described \cite{Grid_aware, preprint_gmm})}
    \label{gmm}
\end{figure}

GMM is a probabilistic and parametric method used for classification, clustering and forecast. In \cite{Grid_aware}, it is used for forecasting the EV charging flexibility, thus applying perfectly to our case study. This method is used for generating $M+1$ scenarios, $M$ for feeding an EV demand forecast to the stochastic DA problem and the last one for the realization. The method, described in Figure \ref{gmm}, consists in several steps. First, a session database is preprocessed according to the day type and the selected features, i.e. arrival time, stay time and energy demand. The number of daily sessions is computed in $M+1$ scenarios through the best univariate GMM using K-fold Cross Validation (CV), scaling the output with the number of Charging Columns (CC). In the meantime, each feature is modeled via univariate, multivariate and mixed GMM with the same K-fold CV approach for selecting the best model for each type. Among these models, the best ones are selected via a combined metric including accuracy, bias and correlation measures, as described extensively in \cite{preprint_gmm}. These models are then used to describe the sessions inside each $M+1$ scenarios. Finally, $M$ scenarios are used to generate a cumulated EV demand probabilistic forecast evenly distributing the charging sessions. For the realization, a measurement database is used to link the sessions of the $(M+1)^{th}$ scenario to real measurements. To do that, the measurement database is filtered according to the arrival time and the closest sessions energy-wise is chosen. In this way, the realization will be composed of real measurements from a real session.

\begin{figure}[t]
    \centering
    \includegraphics[width=\linewidth]{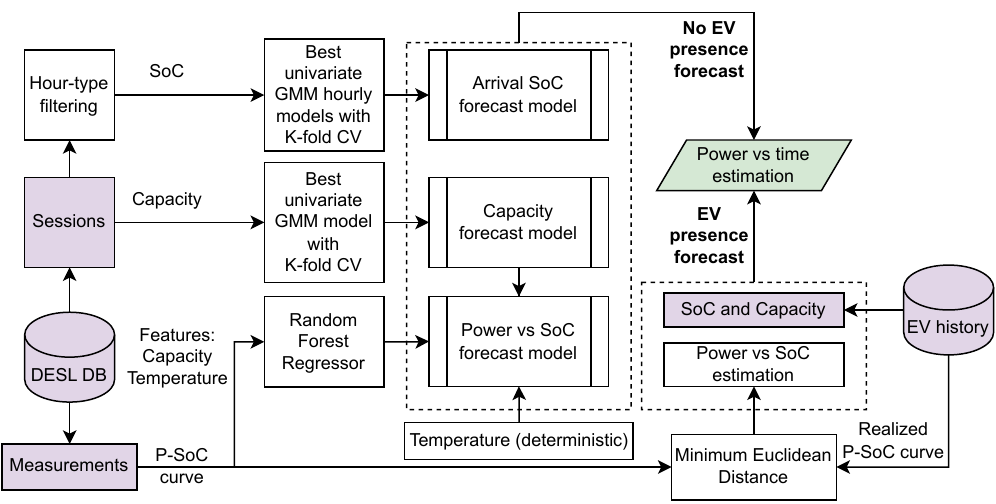}
    \caption{Proposed method for EV intraday forecaster. Two methods are framed - with and without the EV presence - to forecast the power versus time curve of each EV within the booking horizon $\mathcal{T}_{BM}$.}
    \label{intra_ev}
\end{figure}

The comprehensive intraday model for the expected EV demand forecast is described in Figure \ref{intra_ev}. It holds on the assumption that the booking system collects each EV stay time in the intraday horizon and, for connected EV, it reads the historical power versus State of Charge (SoC) curve. Therefore, to estimate the power versus time curve, two different methods are employed according to the EV connection status. Both methods rely on databases deprived from the realization sessions and measurements to avoid biased fitting of the models.  For not connected EV, the capacity is estimated via the best univariate GMM using K-fold CV, previously trained on the session database, to be provided to a Random Forest (RF) regressor, trained on the measurement database, to obtain a full (from 0 to 100\% SoC) power versus SoC curve. To obtain the power versus time curve, a hour-specific GMM (the best univariate one via K-fold CV) is employed to obtain the estimated arrival SoC that is fed to an EV surrogate model to create the required output over the connection time of the EV, available from the booking system. For connected EVs, we assume access to their historical charging profile, i.e., for each past horizon $\mathcal{H}$ we know the SoC, the requested power, and the delivered power. Moreover, we assume to know their battery capacity. As a fallback strategy, we apply the charging model proposed in \cite{epfl_db_analysis} on the same EV database. Therefore, we apply the minimum euclidean distance between the historical P-SoC curve and the capacity, and a carefully sliced measurement database to get the power versus SoC curve estimation. The same EV surrogate model is used to obtain the power versus time curve considering the known current SoC and EV capacity.

\subsubsection{Imbalance price forecast}
We use a naive 1-day persistence for estimating the imbalance prices, corresponding to $\hat{r}_{short,T}$ and $\hat{r}_{long,T}$ of Part I.

\section{Benchmarks} \label{benchmarks}
We propose three benchmark methods to SG-ADMM for the real-time management of the CS. To compare our method with well-established approaches, we propose two different optimization techniques, i.e. a centralized approach (Section \ref{centralized}) and a distributed approach (Section \ref{decentralized}), and an uncontrolled approach (Section \ref{uncontrolled}). SG-ADMM and its benchmarks all use a real-time dispatch described in Section \ref{rtd}.

\subsection{Centralized approach} \label{centralized}
The centralized approach allows the CS controller to impose the setpoints to each EV, therefore requiring full access to all EV information and thus not preserving privacy. This privacy-unaware approach involves the creation of a further communication from each Charging Point (CP) to the CS controller, so that the EMS can optimize each CP power together with the slack and the incentives. We design an hard-to-beat centralized controller having information on the whole set of available information needed from the EV. The formulation of the centralized objective function is as follows:
\begin{equation}
        \operatornamewithlimits{min}_{\mathcal{P}, \Theta, s_L} \sum^N_{i=1} \left (f^*_i(P_i) + \hat{f}_i(P_i) + \theta_i \cdot P_i \cdot \Delta t \right)
\end{equation}
Where, similarly to Eq. (36) of Part I, the goal is to minimize the leader-specific separable objective function ($f^*_i(P_i)$), the follower-specific objective function ($\hat{f}_i(P_i)$) and the revenue loss from incentive provision ($\theta_i \cdot P_i\cdot \Delta t$).

The problem is constrained as follows:
\begin{equation}
    \begin{aligned}
                & s^-_L \leq |s_L^{min}|\text{,} \quad  s^+_L \leq s_L^{max} \text{,} \quad s^-_L \perp s^+_L \\
                &\text{with $s_L^{min}$ and $s_L^{max}$ as in Part I}\\
        & \sum _{i=1}^{N} \frac{P_i}{\eta_{cp}} = C + s^+_L - s^-_L \quad \text{with $C$ as in Part I}\\
        & \sum_{i \in CC} P_i \leq P_{CC} \\
        & \theta_i = min(D, |\delta \cdot \nabla_{P_i} \hat{f}_i(P_i)|)  \quad \text{with $D$ as in Part I}\\
    \end{aligned}
\end{equation}

The problem is formulated as a Mixed Integer Linear Programming where $\hat{f}_i(P_i)$, that is a piecewise function, is written via two positive slack variables indicating the deviation between each EV required and satisfied power. $P^+_i$ represents the positive deviation and $P^-_i$ the negative deviation, therefore:
\begin{equation} \label{}
\begin{aligned}
    & P_i -P_{req,i}  = P^+_i - P^-_i\\
    & \hat{f}_i(P_i) = \beta \cdot (P^-_i)^2 + \gamma \cdot \left(\frac{SF_{P_{req,i}+P^+_i}}{SF_{P_{req.i}}}-1\right)
\end{aligned}
\end{equation}
Where $SF_{P_{req,i}+P^+_i}$ is a highly non-linear function that is approximated through piecewise linearization. This function require from each EV the current SoC ($SoC_i$), the required power ($P_{req, i}$) and the battery capacity ($C_{i}$).

The incentives $\theta_i$ are derived as the minimum value between $D$, obtained in the ID layer, and the absolute value of the derivative with respect to $P_i$ of the individual objective function $\hat{f}_i$. This individual function - and thus its derivative - is a complex piecewise function written as:
\begin{equation}
    \nabla_{P_i}\hat{f}_i(P_i) = -2\cdot P^-_i + \frac{SF'_{P_{req,i}+P^+_i}}{SF_{P_{req.i}}}
\end{equation}
Where $SF'_{P_{req,i}+P^+_i}$ is derived mathematically and approximated via piecewise linearization. 

\subsection{Distributed approach} \label{decentralized}
The distributed approach works with the same principle of the proposed method but without including the game theoretic incentive mechanism design, thus with no leader update. Therefore, the objective function is limited to:
\begin{equation} 
\begin{aligned}
&\hbox{Followers:} \quad P_i^*=\operatornamewithlimits{argmin}_{P_i} \Phi_{i}(\hat{f}(P_i),0) \quad \forall \; 1\leq i \leq N, \\ 
&\hbox{Constraint:}\;\;\; \sum _{i=1}^{N} \frac{P_i}{\eta_{cp}} = C \text{, } \sum_{i \in CC} P_i \leq P_{CC}\\ 
\end{aligned} 
\end{equation}

Where, if the incentive mechanism is not in place, the followers' cost function becomes:
\begin{equation}
    \Phi_{i}(\hat{f}(P_i),\theta^k_i) = L_i(P_i,\lambda) + \hat{f}_i(P_i)\\
\end{equation}

Thus, only the inner loop (ADMM) of the proposed method will be performed, including the sequential follower update ($\theta^k_i=0$), the leader dual update ($\lambda^k$ and $\mu^k_{CC}$) and the inner convergence criterion.

\subsection{Uncontrolled approach} \label{uncontrolled}
Lastly, we compare the proposed method with a greedy and myopic uncontrolled approach, that tries to satisfy the EV exact demand by increasing (or curtailing) the power generation from the grid and the BESS.

\subsection{Real-time dispatch} \label{rtd}
All the real-time methods require a dispatch considering the PV realization, explained in Algorithm 1. The slack is managed by the grid, provided that it remains in the operational constraints, while the PV realization is managed by the BESS. In case BESS saturates to the operational constraints, the surplus of charging or discharging power is managed by the grid, regardless from the operational constraints.

\begin{algorithm} 
\caption{Real-time dispatch}
\begin{algorithmic}[1]
    \State \textbf{Input:} $C$, $s_L$, $P_B$
    \State \textbf{Output:} $P_G$, $P_B$
    \State $P_G = min\left(P_{GC},(C + P_B - P_{PV}^r\cdot\eta_{pv})/\eta_{tr}\right)$
    \State $P_B = P_G \cdot \eta_{tr} + P^r_{PV} \cdot \eta_{pv} - C$
    \If {$P^{ch}_B > C_B \cdot C_{rate}$} 
    \State $\Delta P = P^{ch}_B - C_B \cdot C_{rate}$
    \State $P^{ch}_B = C_B \cdot C_{rate}$
    \State $P_G = P_G - \frac{\Delta P}{\eta_{inv} \cdot \eta_{ch} \cdot \eta_{tr}}$
    \EndIf
    \If {$P^{dh}_B < -C_B \cdot C_{rate}$}
    \State $\Delta P = - C_B \cdot C_{rate} - P^{dh}_B$
    \State $P^{dh}_B = - C_B \cdot C_{rate} $
    \State $P_G = P_G + \Delta P \cdot \eta_{inv} \cdot \eta_{dh} \cdot \eta_{tr}$
    \EndIf
\end{algorithmic}
\end{algorithm}

\section{Results and Discussion} \label{results}
In this section the results are presented and discussed: a daily simulation is firstly analyzed for understanding the overall framework functioning (Section \ref{daily}), followed by a weekly simulation where SG-ADMM is analyzed according to several metrics (Section \ref{weekly}).
\subsection{Daily Simulation} \label{daily}
\begin{figure}[b]
    \centering
    \includegraphics[width=1\linewidth]{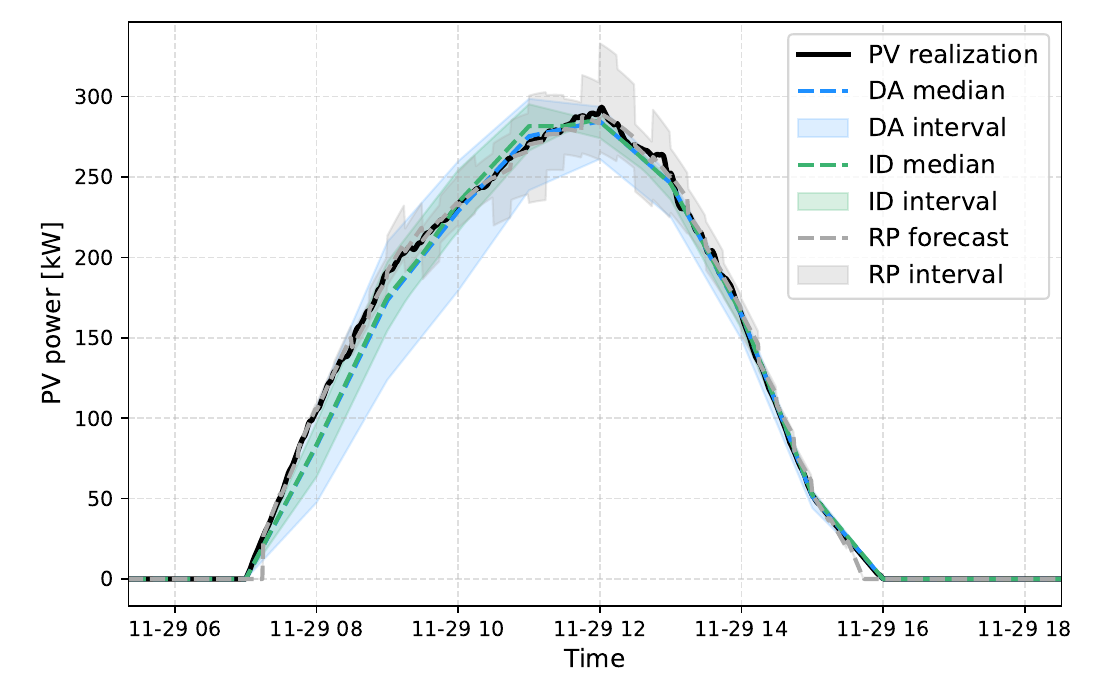}
    \caption{PV forecast and realization for the simulated day: both the median and the interval are shown, together with the realization.}
    \label{pv_for}
\end{figure}

\begin{figure}
    \centering
    \includegraphics[width=1\linewidth]{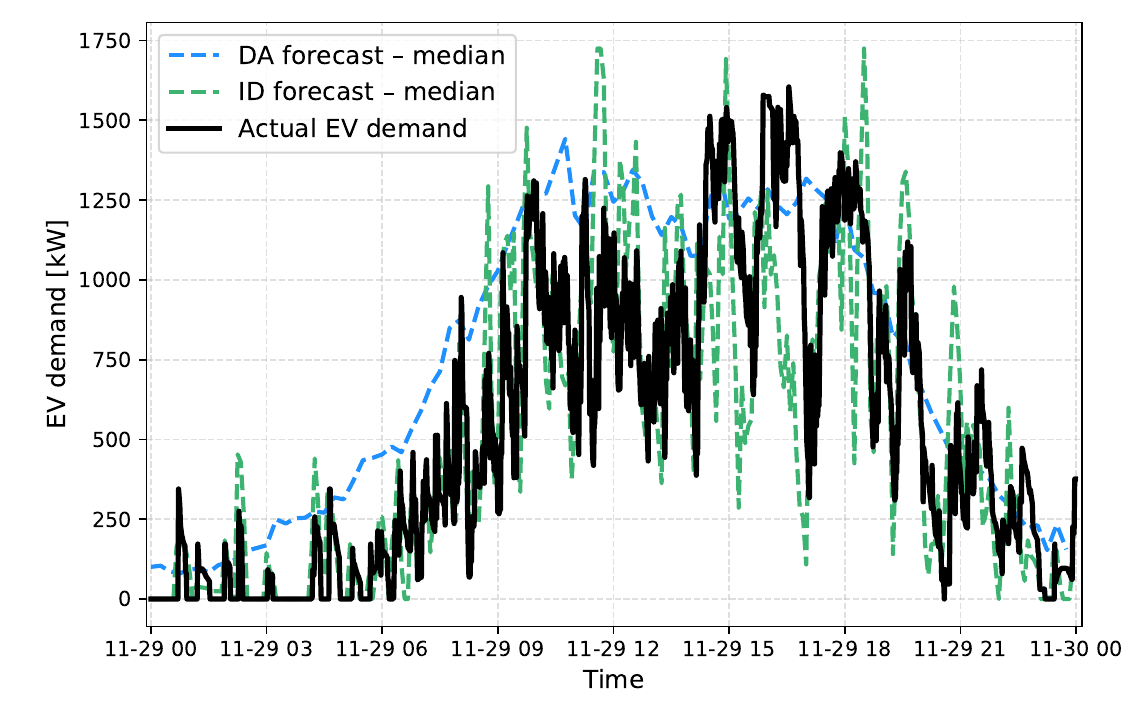}
    \caption{EV forecast and actual demand for the simulated day.}
    \label{ev_for}
\end{figure}

\begin{figure}
    \centering
    \includegraphics[width=1\linewidth]{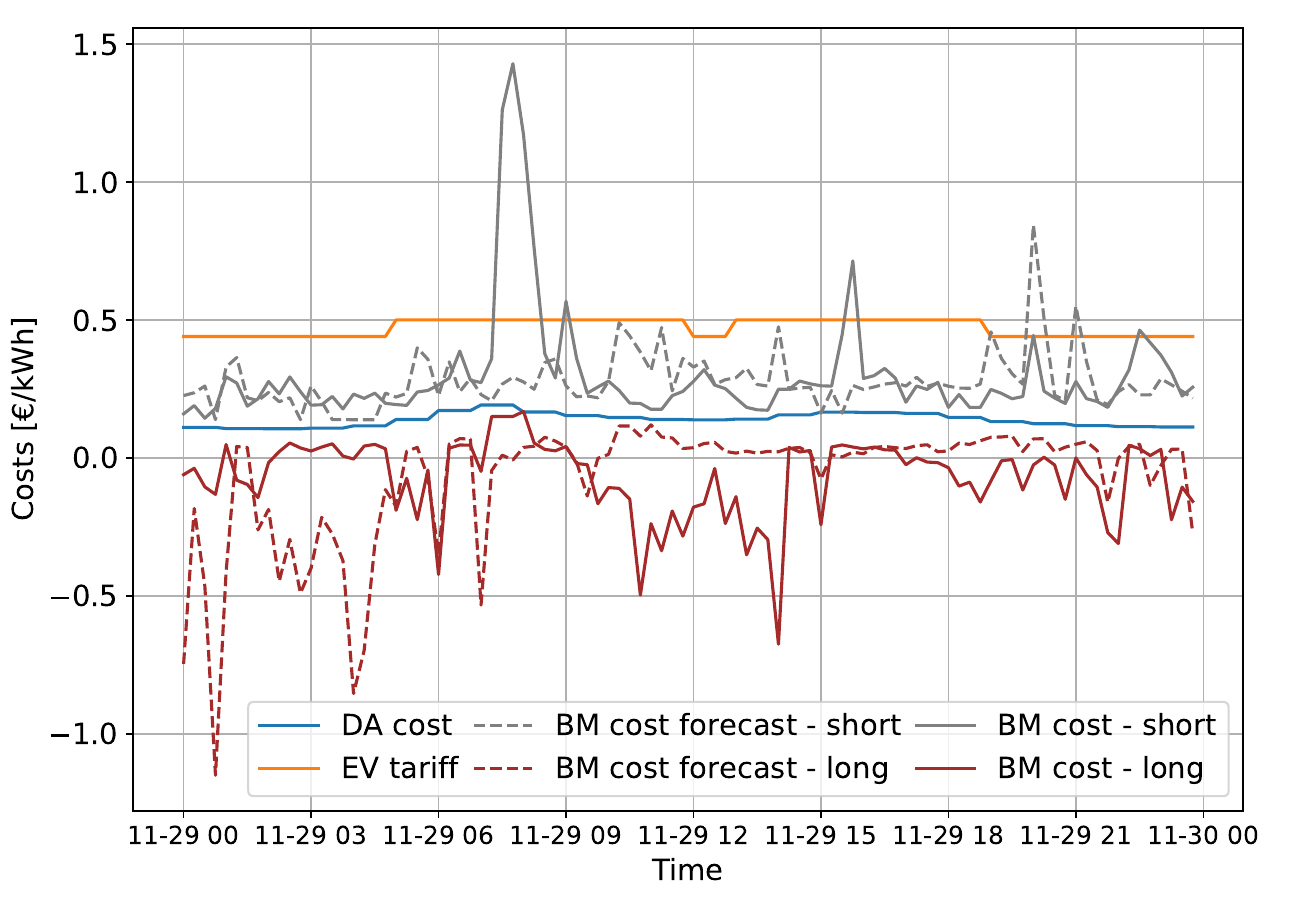}
    \caption{Cost parameters: deterministic (DA cost and EV tariff) and forecasted (BM cost).}
    \label{fig:costs}
\end{figure}

We simulate the $29^{\text{th}}$ of November 2024. Firstly we present the corresponding PV forecasts and realizations in Figure \ref{pv_for}. Similarly, we present the EV forecast and realization in Figure \ref{ev_for}, focusing on the median. In Figure \ref{fig:costs}, we present the economic parameters (DA cost, EV tariff) together with the BM prices (and the corresponding forecasts) for the simulated day. These forecaster will provide the input to the different layers of the proposed EMS.

\begin{figure}[!t]
  \centering
  \subfloat[Daily schedule generated by the DA model.\label{fig:schedule}]{
    \includegraphics[width=\linewidth]{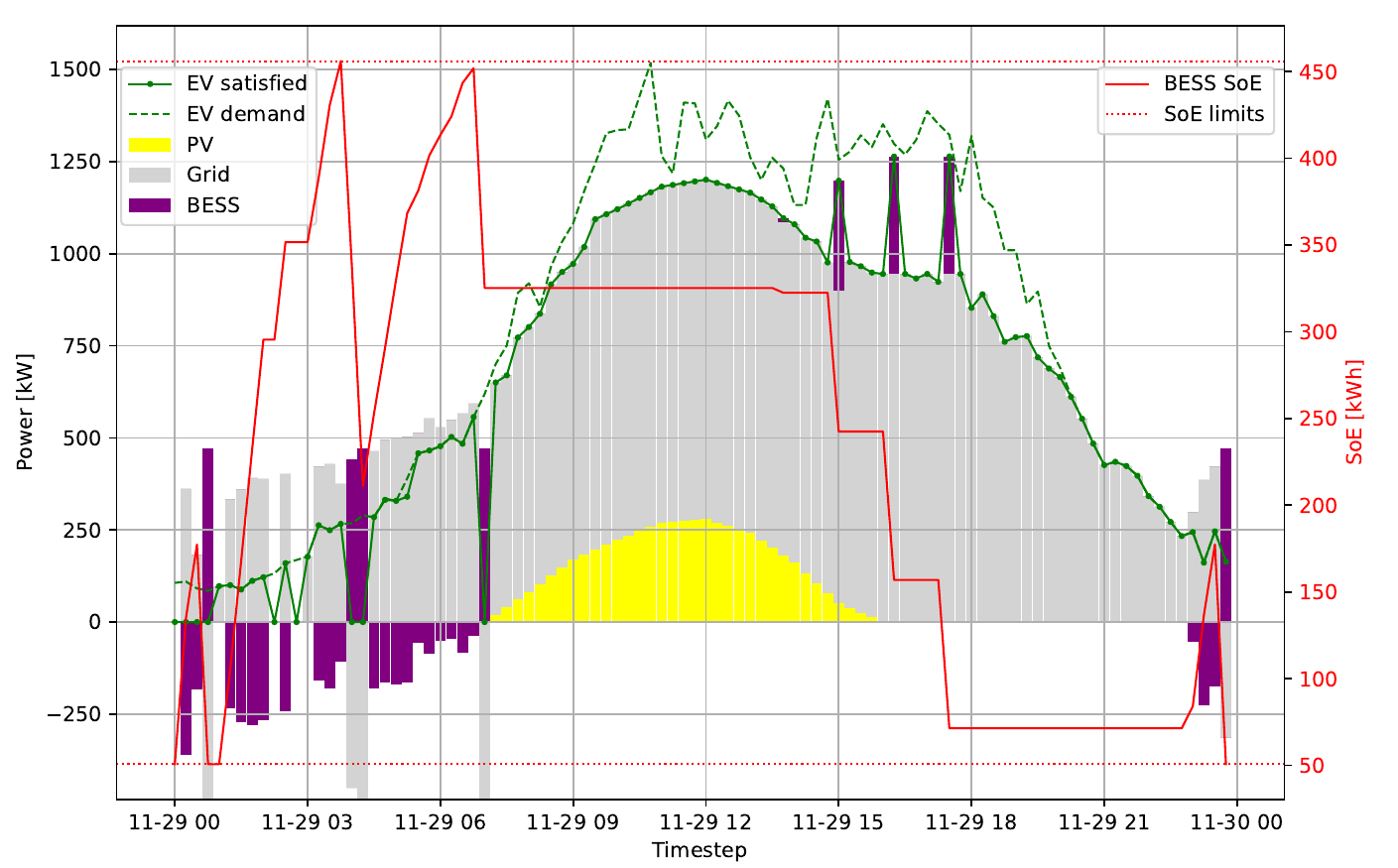}
  }

  \subfloat[Model speculation between the DAM (DP) and the BM (Schedule).\label{fig:speculation}]{
    \includegraphics[width=\linewidth]{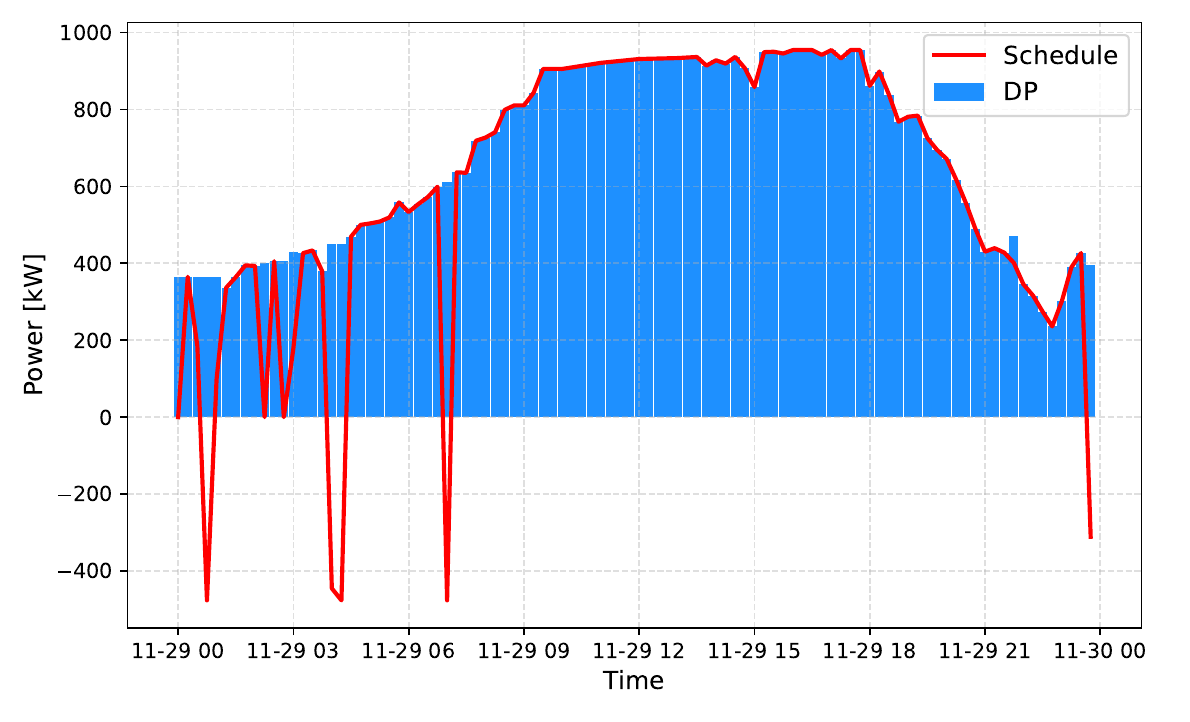}
  }

  \caption{DA model output for the simulated day.} \label{fig:day ahead market}
\end{figure}

\begin{figure*}[t]
    \centering
    \includegraphics[width=1\linewidth]{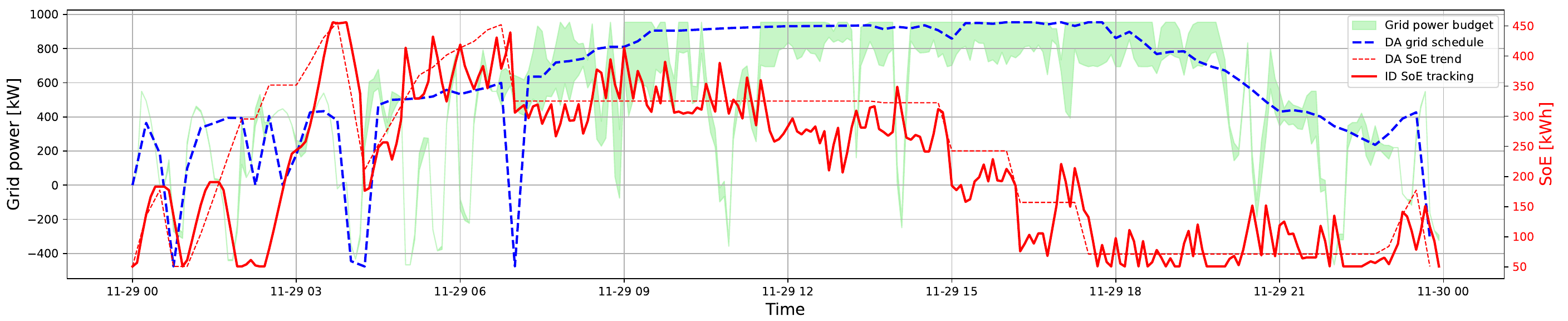}
    \caption{ID model output for the simulated day: grid power budget definition and updated SoE trend. Both are compared to the DA predefined schedule.}
    \label{fig:id}
\end{figure*}

\begin{figure*}[h!]
  \centering

  \subfloat[SG-ADMM\label{fig:sgadmm}]{
    \includegraphics[width=0.49\linewidth, height=0.285\linewidth]{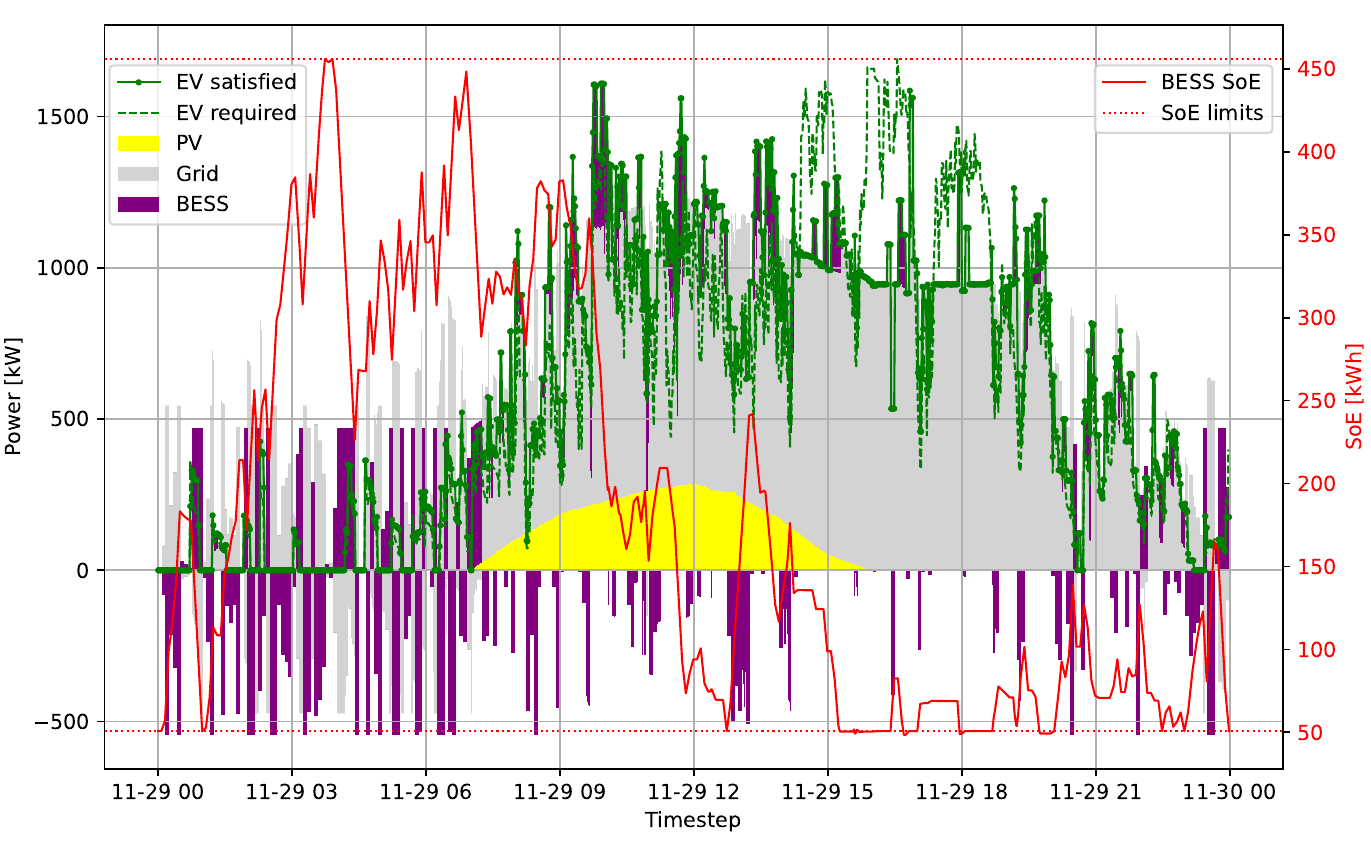}
  }
  \hfill
  \subfloat[ADMM\label{fig:admm}]{
    \includegraphics[width=0.48\linewidth]{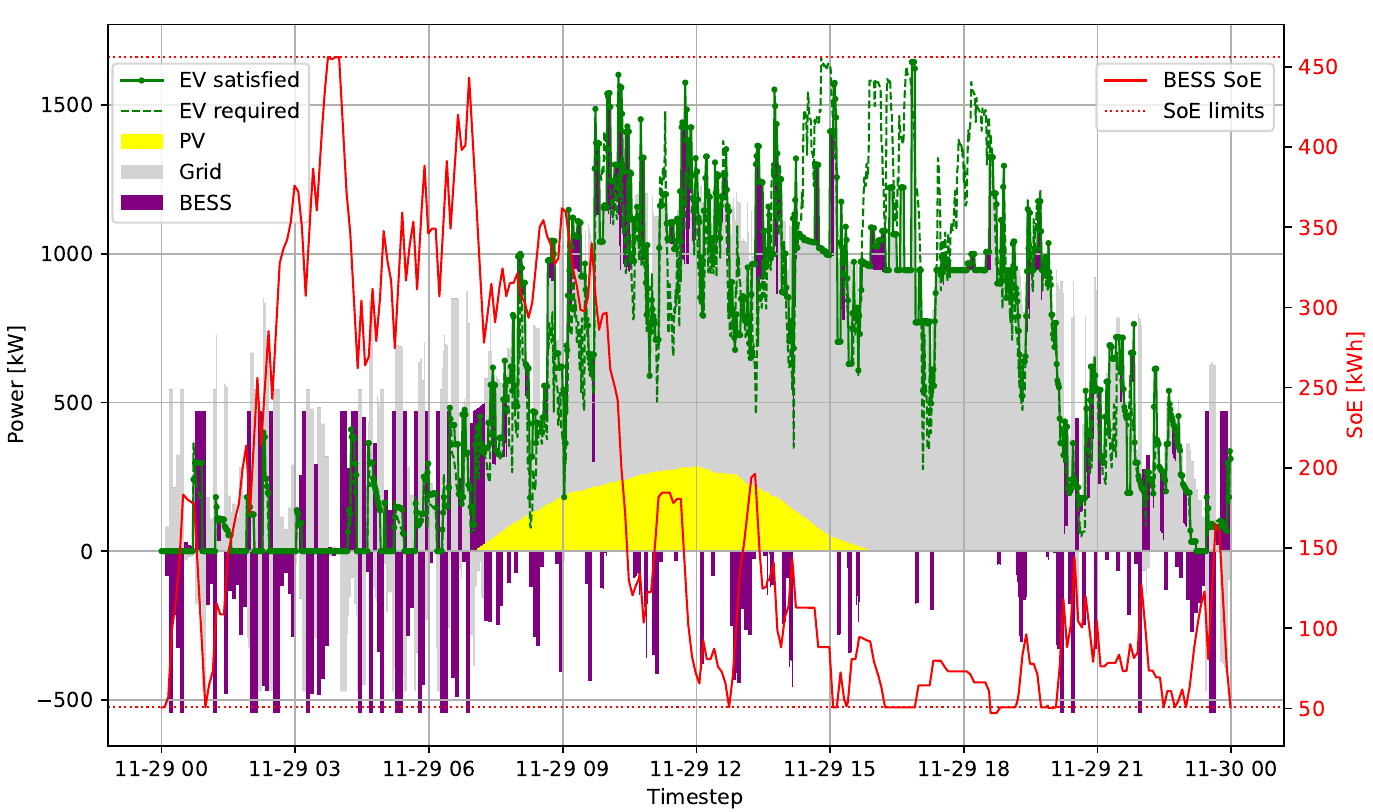}
  }

  \vspace{2mm}

  \subfloat[Centralized\label{fig:central}]{
    \includegraphics[width=0.485\linewidth]{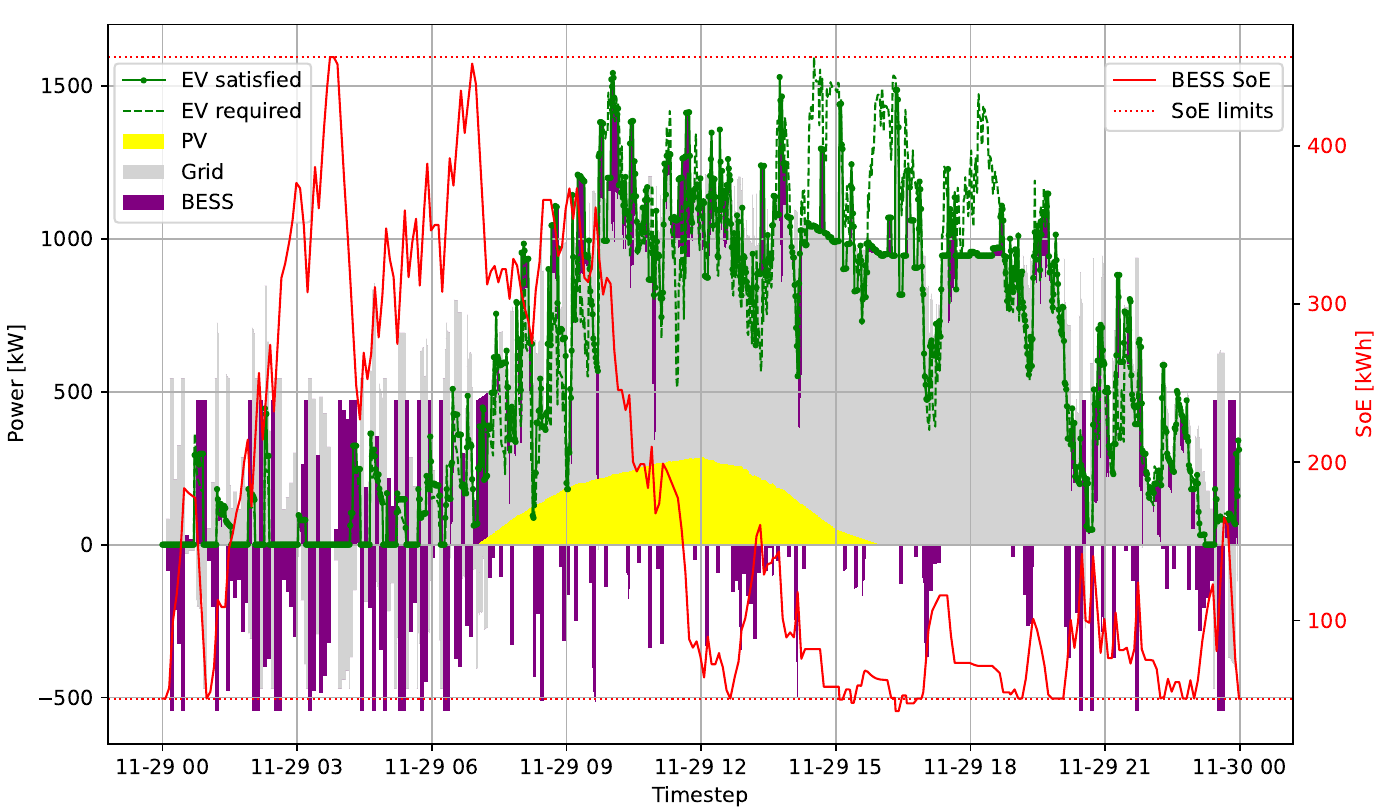}
  }
  \hfill
  \subfloat[Uncontrolled\label{fig:uncontrolled}]{
    \includegraphics[width=0.485\linewidth]{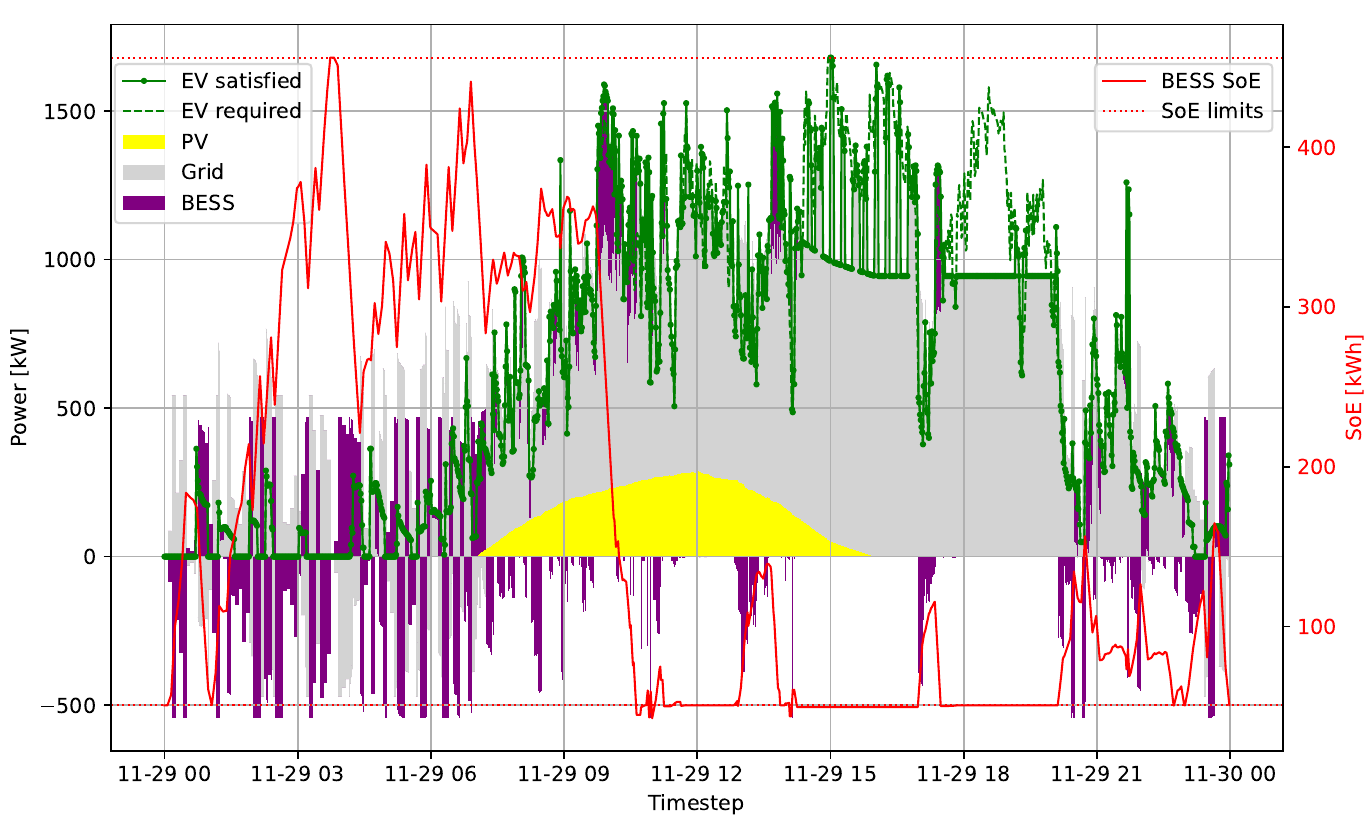}
  }

  \caption{Real-time output}
  \label{fig:rt}
\end{figure*}

At midnight, the DA model generates a schedule considering both the DAM and the BM. In Figure \ref{fig:day ahead market} we show the output of the DA model: in Figure \ref{fig:schedule} we show the schedule, while in Figure \ref{fig:speculation} we show the extent of the CS speculation comparing the schedule to the DP. The schedule presented in Figure \ref{fig:schedule} makes clear how the stationary battery is planned to be used, with almost two full equivalent cycles. Between 12 a.m. and 7 a.m., the BESS is mainly used to speculate rather than satisfying the limited EV demand expected for that period. In fact, low expected long prices during the early hours of the day (see red dashed line in Figure \ref{fig:costs}) steer the system to deviate from the DP submitted to the DAM, considering that the ID model will be able to refine the choice according to the real BM prices. This would be the case as long prices exhibit a low forecast accuracy until 6 a.m., as it will be shown later. Moreover, not only the BESS will be used to speculate, but in some cases (when the EV satisfied - green bold line - is zero and the EV demand - green dashed line - is not, in Figure \ref{ev_for}) the overall system considers more convenient to speculate rather than satisfying the EV demand at all. Between 7 a.m. and 10 a.m., the EV satisfied always falls short to the EV demand, even in case in which the grid alone would have been able to sustain the power request. This is because the chance-constrained proposed formulation controls EV and BESS, thus limiting the grid withdrawal in a robust way considering PV underestimation error. Around 10 a.m., where the PV error is less impacting, the model starts satisfying all the EV demand just by using the grid. Indeed, the BESS is not used to cover for EV peaks by the day-ahead model, except between 3 p.m. and 6 p.m. where the discharge helps in getting additional revenues in peak tariff (see orange bold line in Figure \ref{fig:costs}). From 8 p.m. onward, the EV demand is completely satisfied, while after 10 p.m. an additional charge and discharge process is envisioned to speculate on the expected long price. This behavior is reflected also on the comparison between the DP communicated to the DAM and the internal CS schedule in Figure \ref{fig:speculation}, where the comparison clarifies the trend to speculate only through long prices, knowingly causing negative imbalances.

Then, the ID model is run every 15 minutes with 5 minutes granularity. We show in Figure \ref{fig:id} the grid power budget allocation - compared to the grid schedule - and the BESS State of Energy (SoE) - compared to the DA one. For each ID update, the graph only contains the most updated one. The results highlight the corrections that the model performs once aware of the BM prices and considering the updated PV and EV forecasts. As a response to this new inputs, the ID model generates a more variable grid power budget profile, that better follows the expected EV demand variability that is not incorporated in the day-ahead forecast (as can be seen from Figure \ref{ev_for}). Moreover, in the ID update we consider the imbalance prices as known, therefore shaping the grid power budget as can be seen in the low long price between 9 a.m. and 3 p.m. For what concerns the battery scheduling, the SoE is overall well tracked except between 2 a.m. and 4 a.m. where a much higher long price impacts the initial scheduling. The other deviations in SoE tracking are linked to the different forecast on EV demand; when DA forecast estimates a higher cumulative demand than the ID forecast, the SoE remains above the reference one and vice versa.

In Figure \ref{fig:rt} we show, for each simulated method, the real-time control in terms of power flows. Firstly, the four methods all have similar behavior in managing the realization of EV demand and PV production with respect to the ID output. The main difference between the Uncontrolled approach (Figure \ref{fig:uncontrolled}) and the optimized methods is the flexibility that the other methods exploits by increasing or decreasing the EV satisfied power with respect to the required one. Another difference is the extent of BESS charging around 1 p.m.: in fact in 30 minutes, SG-ADMM (Figure \ref{fig:sgadmm}) stores around 250 kWh (more than the 100 kWh for ADMM, centralized and uncontrolled) to restore the SoE tracking. In fact, while until 7 a.m. the low EV demand allows to follow precisely the predefined BESS schedule, the EV realization deviates from the forecasted one thus requiring the real-time control to use the BESS as a buffer. This explains why, comparing Figure \ref{fig:id} and \ref{fig:rt}, the SoE reaches the lower limit sooner than 3 p.m. Generally, SG-ADMM and centralized (Figure \ref{fig:sgadmm} and Figure \ref{fig:central}, respectively) manage both the BESS and the grid exchange in a less variable way. In fact, ADMM (Figure \ref{fig:admm}) shows a larger number of charge–discharge cycles and a more unstable BESS trajectory, indicating a more reactive tendency towards forecast errors, and uncontrolled (Figure \ref{fig:uncontrolled}) displays the most aggressive use of the BESS and frequent grid constraint violation.

\begin{figure}[!t]
  \centering

  \subfloat[Negative Slacks\label{fig:slack-}]{
    \includegraphics[width=\linewidth]{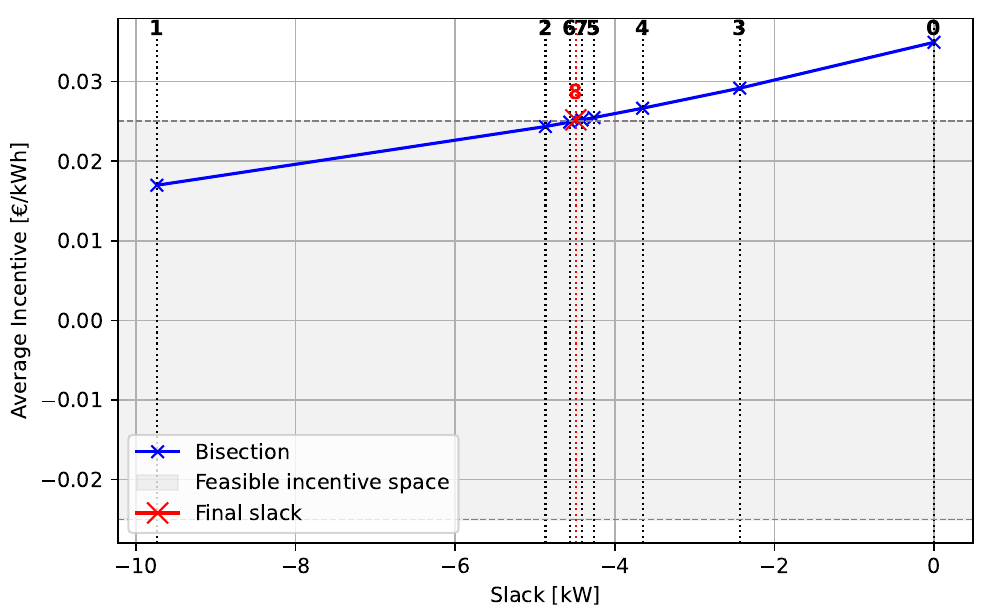}
  }

  \subfloat[Positive slacks\label{fig:slack+}]{
    \includegraphics[width=\linewidth]{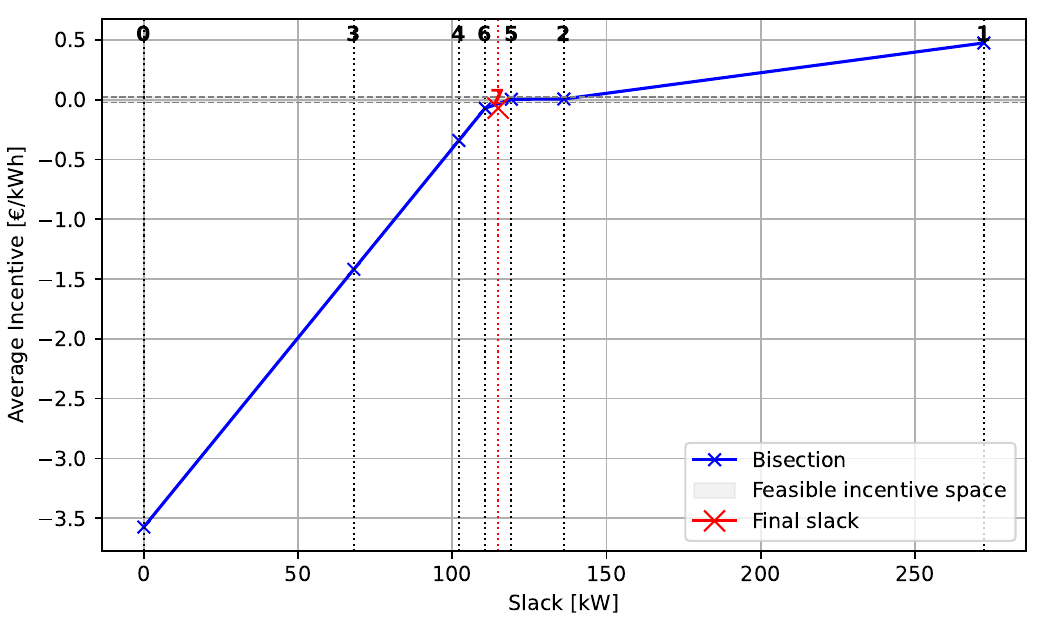}
  }

  \caption{Slack bisection method}
  \label{fig:slack}
\end{figure}

We present in Figure \ref{fig:slack} the functioning of the slack bisection method proposed in Figure 3 from Part I. Our proposed method works with a inner and outer loop, where the inner loop is the ADMM update and the outer loop the Stackelberg game. In the outer loop, the Stackelberg equilibrium is obtained by a bisection method that aims at finding the lowest slack that corresponds to a feasible incentive. We show in Figure \ref{fig:slack-} and in Figure \ref{fig:slack+} how this tweaking to the original SG-ADMM works, respectively for negative and positive slack ranges. As it can be seen, the bisection method originally evaluates the extremes of the slack range (case 0 and 1 on the figures), and it proceeds bisecting the slack space to reach the lowest slack corresponding to a feasible incentive. Each bisection corresponds to a outer loop iteration, thus for the two examples it takes respectively 8 and 7 SG iteration to converge.

\subsection{Weekly Simulation} \label{weekly}
\begin{figure}
    \centering
    \includegraphics[width=1\linewidth]{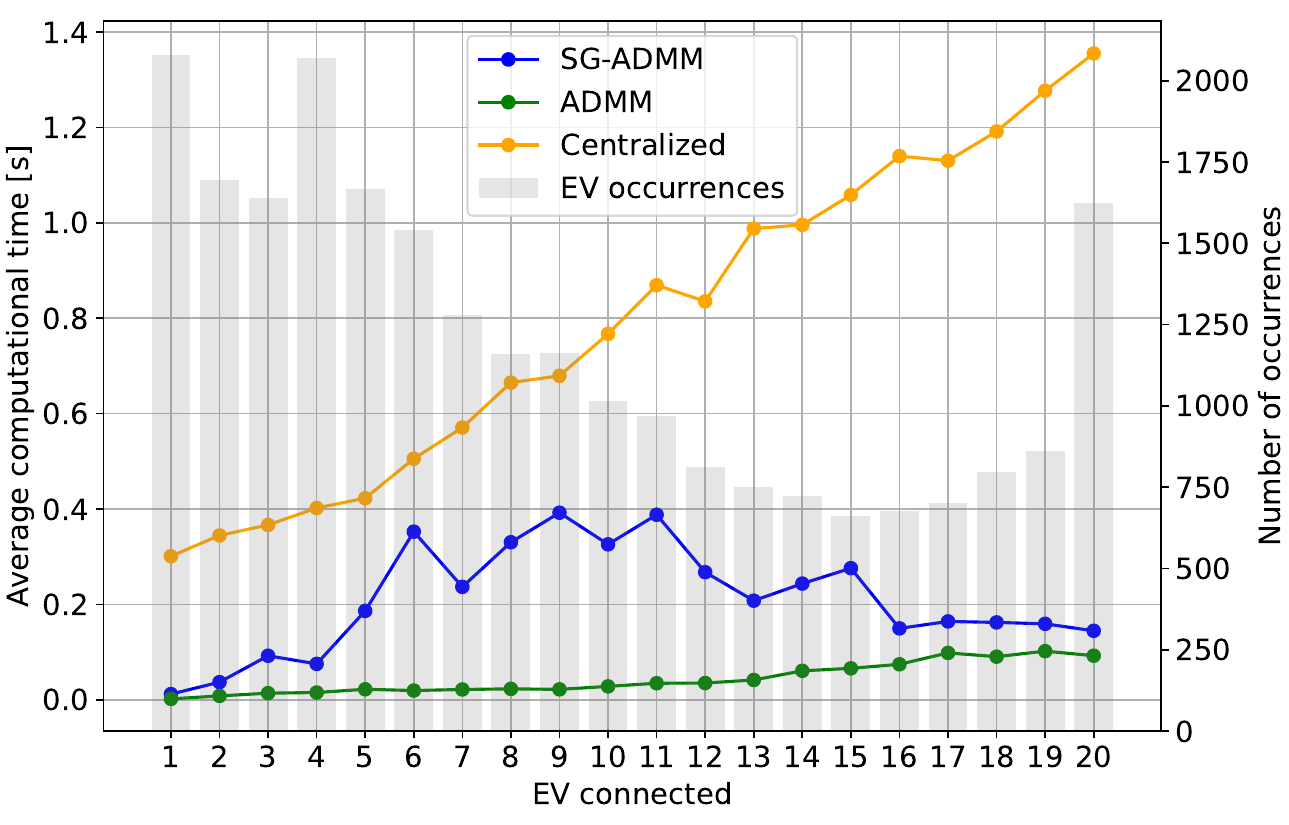}
    \caption{Computational time}
    \label{fig:comp1}
\end{figure}

We simulate the week from the $29^{\text{th}}$ of November to the $5^{\text{th}}$ of December 2024. We present here the results of the comparison between the proposed model and the benchmark models in terms of several metrics, starting from computational time. In Figure \ref{fig:comp1}, we show the average computational time for each number of EV connected to the CS. The simulation have been performed on a PC with 13th Gen Intel\textsuperscript{\textregistered} Core\texttrademark \,i7-13700 @ 2.10 GHz. In the background, we show the occurrence - over a week - of that number of EVs being simultaneously connected. The graph shows that, for each number of EV connection, SG-ADMM lies between a lower bound and a upper bound, i.e. ADMM and Centralized approaches. In particular, ADMM is steadily increasing, while SG-ADMM experiences a maximum between 9 and 11 EVs, after which the computational time decreases and remain stable after 16 EVs. This can be explained considering the case study, where each of the 10 CC has 2 CPs; this means that until 10 EV, finding the Stackelberg equilibrium takes increasingly more time, that remains similar for medium penetration of EVs but decreases as the Stackelberg equilibrium becomes more trivial to find. On the other hand, the Centralized approach has a linear increase of the computational time, as the dimensionality of the decision vector and the number of constraints expand linearly, reaching 1.3 seconds for 20 EVs (against around 0.2 seconds for SG-ADMM and 0.1 seconds for ADMM). These results show us the superior performance of the decentralized model in the scalability of the algorithm, especially if envisaging a real-time control of 1 second, where the Centralized approach couldn't be applied safely.

\begin{figure*}[t]
    \centering
    \includegraphics[width=1\linewidth]{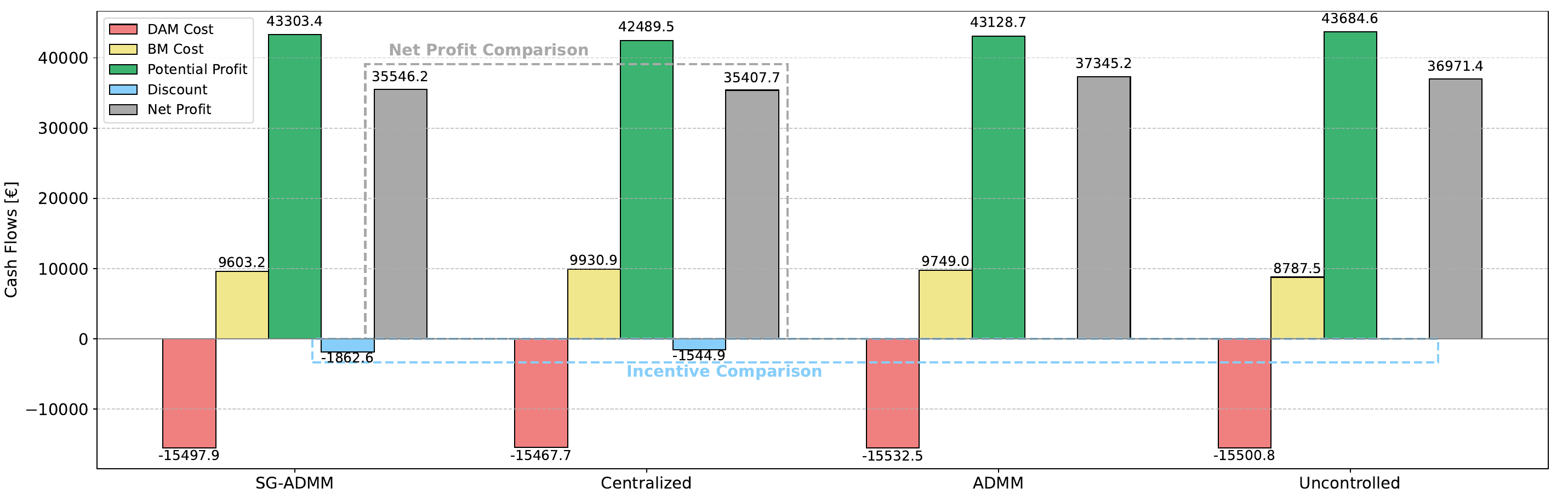}
    \caption{Cash flows comparison of the four methods over a week: the focus is on the net profit comparison and the incentives provided.}
    \label{fig:profit}
\end{figure*}

\begin{figure}[t]
    \centering
    \includegraphics[width=1\linewidth]{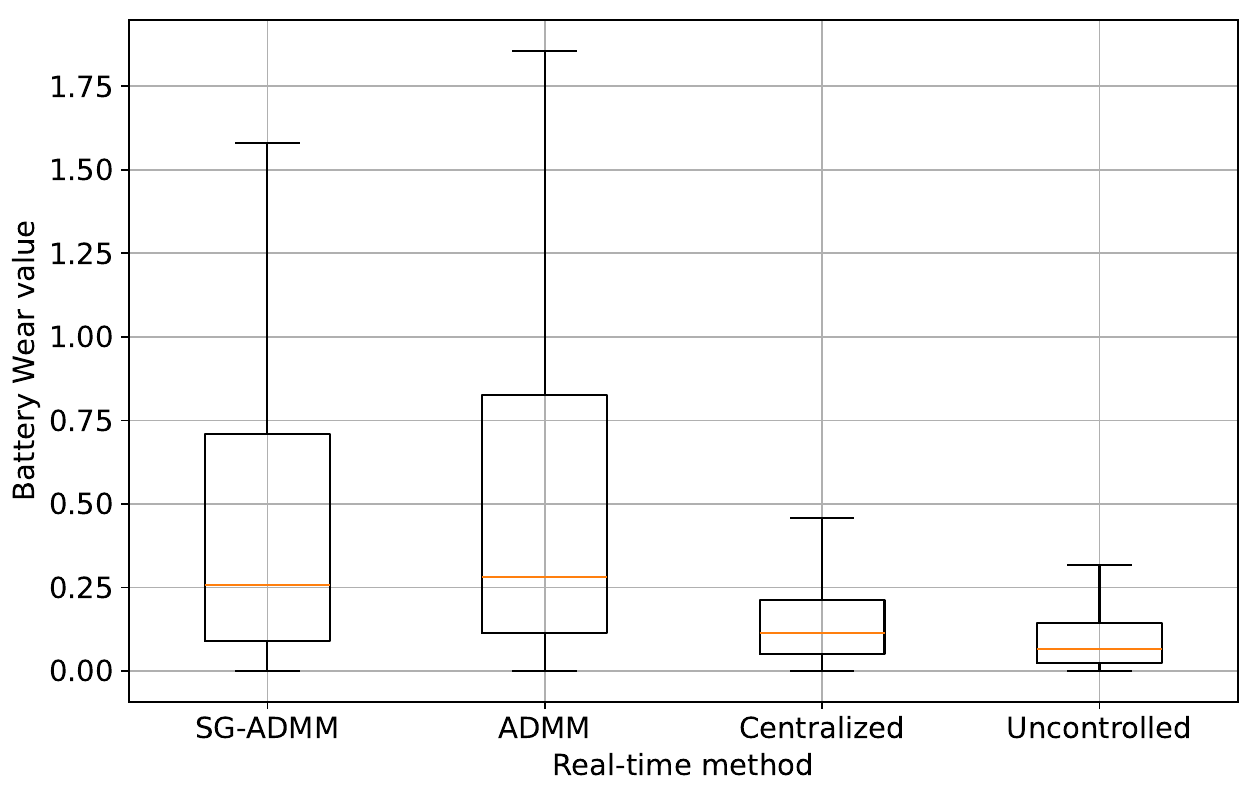}
    \caption{Battery Wear.}
    \label{fig:comp2}
\end{figure}

\begin{figure}[t]
    \centering
    \includegraphics[width=1\linewidth]{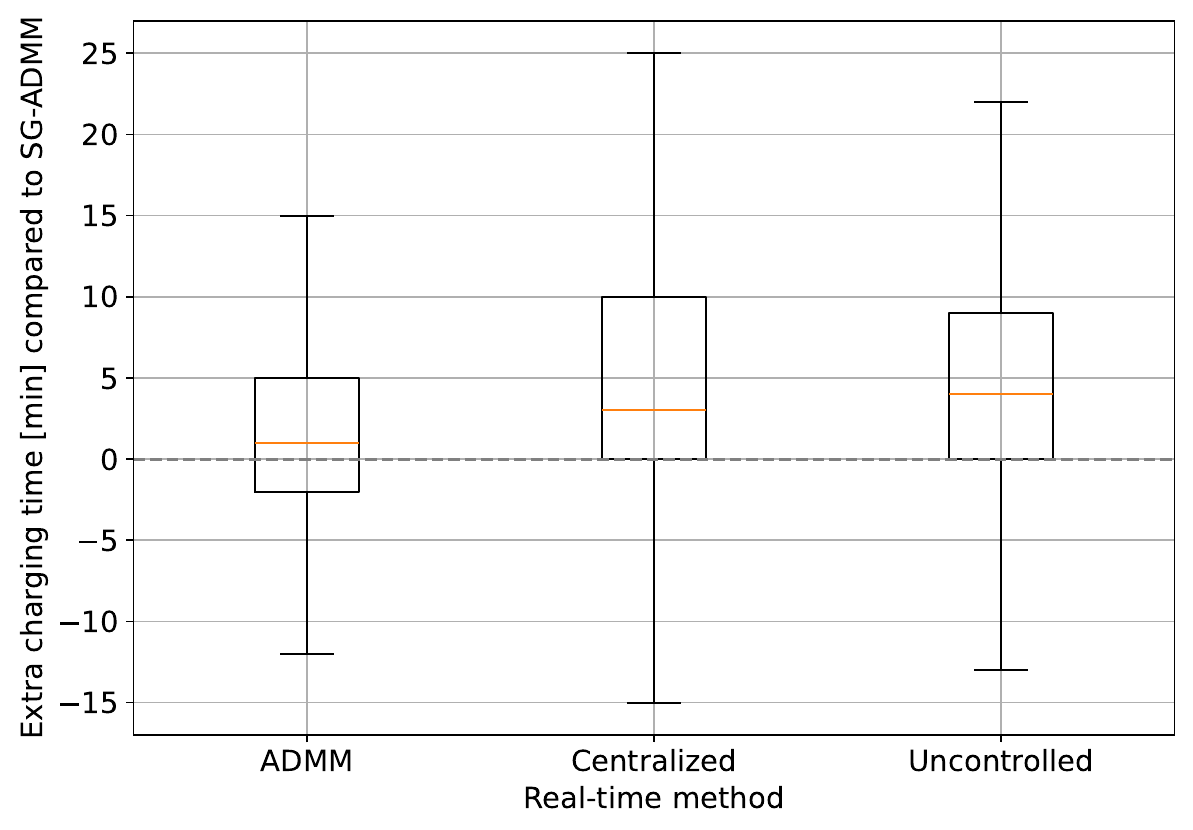}
    \caption{Additional charging time using a different method with respect to the proposed SG-ADMM.}
    \label{fig:chtime}
\end{figure}

\begin{figure}[!t]
  \centering

  \subfloat[Fairness distribution: high value corresponds to lower fairness. \label{fig:f1}]{
    \includegraphics[width=\linewidth]{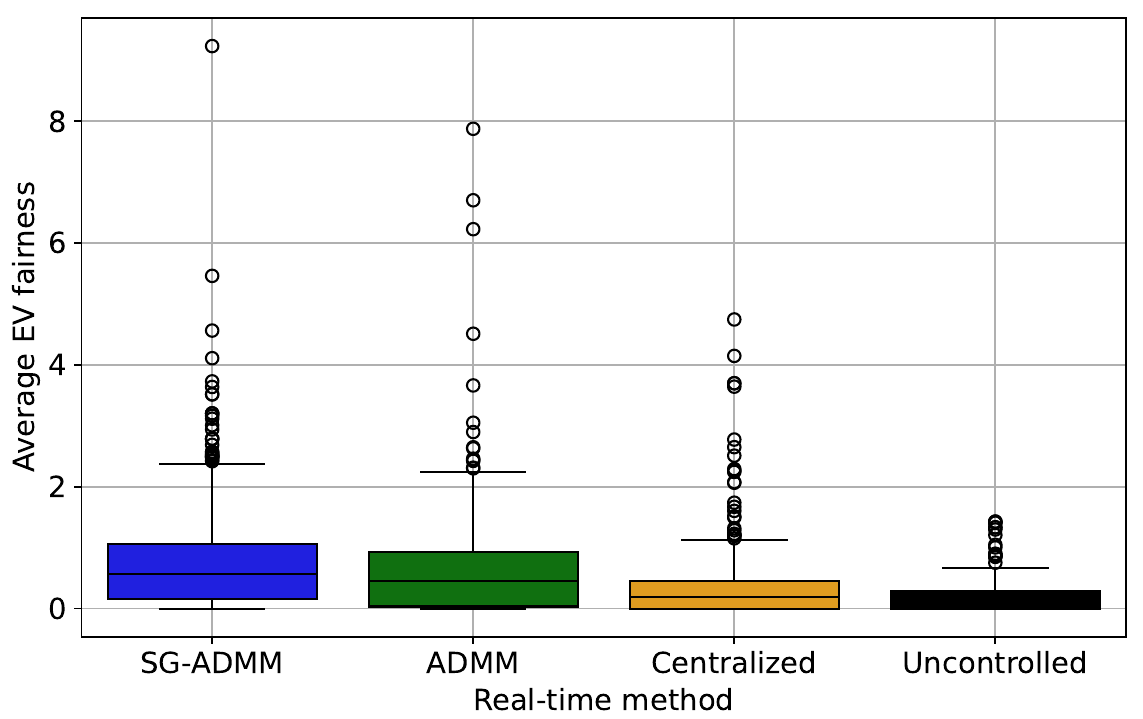}
  }

  \subfloat[Gini index represents how much a distribution is unequal.\label{fig:f2}]{
    \includegraphics[width=\linewidth]{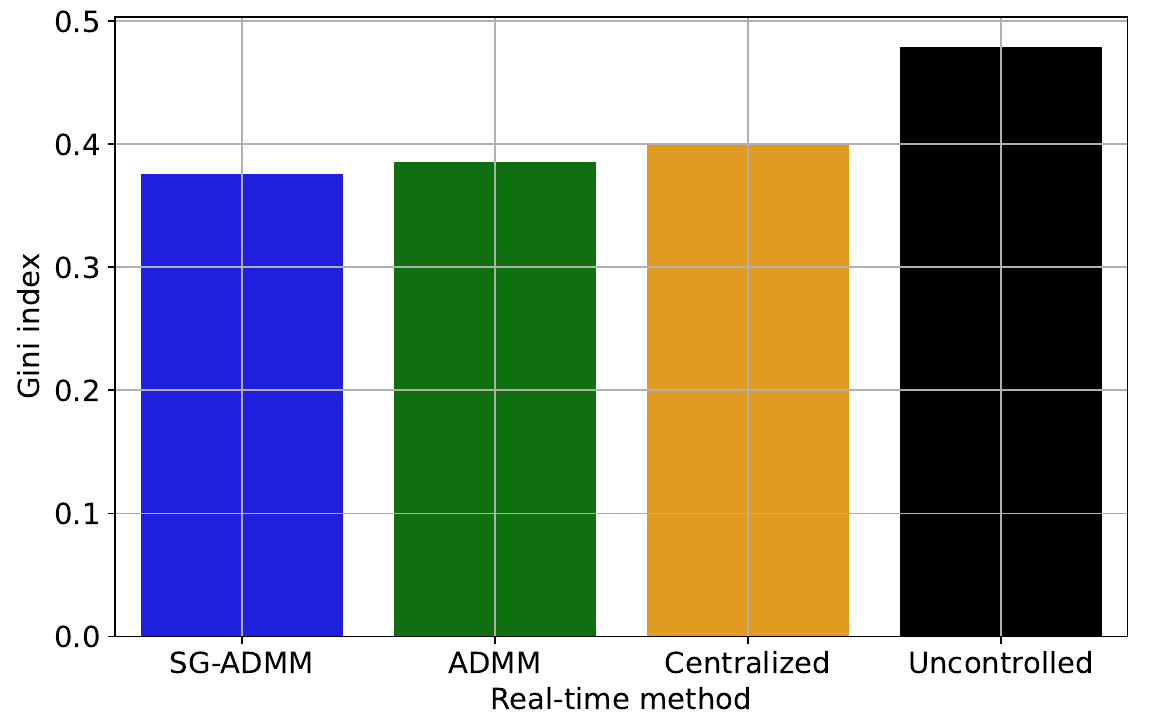}
  }

  \caption{EV fairness distribution analysis.}
  \label{fig:fairness}
\end{figure}

We also evaluate our proposed method in terms of profit and social welfare maximization. In order to do that, in Figure \ref{fig:profit} we show the breakdown of the net profit for the four tested models, focusing on two aspects. Firstly, the net profit of the models that design an incentive (SG-ADMM versus Centralized). Secondly, the incentives provided by each model. The net profit is broken down in DAM cost, BM cost (positive as the CS participates actively in the BM), the potential profit (no costs and no incentives), the incentives and the net profit. When comparing net profit, only the two approaches that incorporate an incentive mechanism, namely SG-ADMM and the centralized method, can be considered. SG-ADMM yields a better result in terms of net profit even with a higher provision of incentives. This is possible due to a higher amount of potential profit that offsets the incentives difference (and also the BM cost that is managed better by the Centralized). This results in a 140 \texteuro \, improvement for the SG-ADMM despite a 320 \texteuro \,increase in incentives. Obviously the methods without incentive yield higher net profit, but these methods offer no incentive despite using the EVs flexibility, thus affecting the social welfare maximization.

For what concerns battery wear, Figure \ref{fig:comp2} shows the distribution of battery wear - computed as in \cite{Rudnik} - in each model. For this metric, the SG-ADMM holds similar performance with respect to ADMM even if our proposed model manages to reduce the extreme battery wear cases (as it can be seen from the upper part of the distribution). Centralized and Uncontrolled obtains better result (Uncontrolled gives inherently smoother power profiles). On the other hand, this increased battery wear means that often the upward flexibility from the EVs is used. This is attested by distribution of the extra charging time with respect to the proposed method, shown in Figure \ref{fig:chtime}. The boxplots shows how the median extra time is always positive and how the 25\%-75\% quantile range lies almost entirely on the positive quadrant, especially for Centralized and Uncontrolled methods. As the proposed method requires a fine-tuning that led to a preference of charging time reduction over battery wear, it is possible to tune the model so that a more battery-aware strategy is implemented.

EV fairness is evaluated considering the absolute per unit deviation with respect to the demand. The result of this evaluation are shown in Figure \ref{fig:fairness}, where Figure \ref{fig:f1} part shows the distribution of the fairness among EVs for each model, while Figure \ref{fig:f2} shows the corresponding Gini index to show the extent of the inequality among the fairness itself. This index is calculated as follows for values sorted in ascending order, where $n$ is the number of EVs and $x$ is the EV average fairness, $G = \frac{2 \sum_{i=1}^{n} i x_i}{n \sum_{i=1}^{n} x_i} - \frac{n+1}{n}$. While the SG-ADMM has a lower fairness, its Gini index is the lowest one. This means that the SG-ADMM exploits more EV flexibilty, resulting in a lower fairness, but in a more uniform way. This ensures that, despite more EVs will not follow the required demand, this process is done in a more democratic way among all the EVs while obtaining highest incentives than the other approach.


\section{Conclusion} \label{concl}
In Part II of this two-part paper, the SG-ADMM real-time control of an EVCS proposed in the first part is contextualized and tested. To validate this proposed real-time approach, we include it in a multi-layered EMS managing the DA and ID scheduling, comparing it with three different benchmark methods in a weekly simulation. To this end, the work introduces a case study and the forecasting modules required by the EMS, with simplified PV forecasting for inter-layer consistency and an adaptive data-driven EV forecasting approach.

The SG-ADMM results demonstrate the applicability of such method, especially in scalability and economic terms, obtaining a lower computational time and a slightly higher net profit - despite the higher incentives - with respect to the Centralized approach. The comparison with the other two methods, ADMM and Uncontrolled, highlights the impact of exploiting flexibility without monetizing it through an incentive to the EV users. Applying a game-theoretic decentralized approach is therefore feasible, computationally efficient and economically sound. In a context where Leader and Followers have different objectives, framing their relationship with a Stackelberg game that trade off the coupling constraint and the incentive has been successful, as the incentive scheme for the EV flexibility exploitation is counterbalanced by the commitment power inherently held by the Leader in a SG. 

This makes the proposed approach particularly attractive for Charging Point Operators (CPO) that are looking for methods that scale with the number of EVs, ensured by the decentralization of the problem, and that incentivize the EVs' flexibility while considering the CPO's own operational and economic objectives, guaranteed by the game-theoretic formulation. Integrating this approach would also attract EV owners that value privacy and fairness, allow responsibilities toward the grid operators to be distributed among the agents, and provide regulators with a concrete example of an incentive-compatible framework for EV-level flexibility exploitation.

The main limitation of the proposed method consists in the required fine-tuning and the amount of hyperparameters to tune. Future works should focus on generalizing the method, reducing the hyperparameters to tune. Moreover, a 1-minute resolution represents a good trade-off between computational time and realistic results, but the model would benefit from an analysis on a 1-second scale, both to highlight the computational time advantage and to demonstrate its applicability also to real testbeds. 

\bibliographystyle{IEEEtran}
\bibliography{literature}

\end{document}